\documentclass{ws-procs975x65}

\begin{document}
\title{THE EFFACING PRINCIPLE IN THE POST-NEWTONIAN CELESTIAL MECHANICS}
\author{Sergei Kopeikin}
\address{Department of Physics and Astronomy, University of Missouri, Columbia, MO 65211, USA\\\email{kopeikins@missouri.edu}}
\author{Igor Vlasov}
\address{Department of Physics, University of Guelph, Guelph, Ontario, N1G 2W1, Canada\\\email{ivlasov@physics.uoguelph.ca}}

\begin{abstract}
First post-Newtonian (PN) approximation of the scalar-tensor theory of gravity is used to discuss the effacing principle in N-body system, that is dependence of equations of motion of spherically-symmetric bodies comprising the system on their internal structure. We demonstrate that the effacing principle is violated by terms which are proportional to the second order rotational moment of inertia of each body coupled with $\beta-1$, where $\beta$ is the measure of non-linearity of gravitational field. In case of general relativity, where $\beta=1$, the effacing principle is violated by terms being proportional to the rotational moment of inertia of the forth order. For systems made of neutron stars (NS) and/or black holes (BH) these terms contribute to the orbital equations of motion at the level of the third and fifth PN approximation respectively.  
\end{abstract}
\bodymatter
It is well-known that in the Newtonian physics as well as in general relativity the external gravitational field of an isolated body having non-rotating, spherically-symmetric distribution of mass, does not depend on the specific internal structure of the body, and is completely determined by a single parameter that is mass of the body. This property of the gravitational field is called the effacing principle \cite{dam}. Effectively, the gravitational field of the spherical body is equivalent to the field of a point-like mass located at the center of mass of the body.

When several bodies form a self-gravitating system they interact to each other and disturb the interior distribution of matter via tidal field. In the Newtonian physics, this disturbance induces body's ellipticity, and leads to appearance of higher multipole moments of the gravitational field of the body describing violation of the effacing principle in the N-body system. Violation of the effacing principle makes the equations of motion of the bodies different from those of the point-like masses. In the Newtonian physics and for a given precision of calculation of equations of motion of celestial bodies one can postpone the violation of the effacing principle by making the characteristic distance between the bodies large enough, thus, reducing the tidally-induced multipole moments to negligible order \cite{kop85}. Indeed, the tidally-induced orbital force \cite{alex} 
\begin{equation}\label{1}  
F_{\rm tide}\simeq \kappa_{\rm tide}\left(\frac{v_{\rm e}}{v_{\rm s}}\right)^2\left(\frac{L}{R}\right)^5 F_N\;,
\end{equation}
where $F_N=GM^2/R^2$ is the Newtonian gravity force for a point-like mass, M and $L$ are characteristic mass and size of the bodies, R is the average distance between the bodies, $G$ is the universal gravitational constant, $v_{\rm e}$ is the body's escape velocity, $v_{\rm s}$ is the speed of sound inside the body's interior, and $\kappa_{\rm tide}$ is a numerical factor depending on the internal distribution of density. Decreasing the ratio $L/R$ can make $F_{\rm tide}\ll F_N$.

The problem of generation of gravitational waves by coalescing NS/BH binaries makes it important to study the problem of violation of the effacing principle in general relativity and alternative theories of gravity. We have used the scalar-tensor theory of gravity to explore this problem in the first PN approximation \cite{kv}. We assume that each body of the N-body system has the center of spherical symmetry 
located at the center of mass of the body that coincides with the origin of the local coordinates associated
with this body. It means that all functions characterizing internal structure of the body have spherically-symmetric
distribution in the local coordinates. We also assume that each body rotates rigidly around its center-of-mass. The rotational deformation leads to the orbital force \cite{alex}
\begin{equation}\label{2}  
F_{\rm rot}\simeq \kappa_{\rm rot}\left(\frac{v_{\rm r}}{v_{\rm s}}\right)^2\left(\frac{L}{R}\right)^5 F_N\;,
\end{equation}
where $v_{\rm r}\simeq\omega L$ is the linear velocity of the body's rotation, $\omega$ is the angular rotational frequency, and $\kappa_{\rm rot}$ is a numerical factor depending on the internal distribution of density. Making $L/R$ sufficiently small one can neglect $F_{\rm rot}$. 

In the first PN approximation of the scalar-tensor theory the orbital equations of motion are \cite{kv}
\begin{equation}\label{3}
M_B\vec{a}_B=\vec{F}_N+\vec{F}_{\rm tide}+\vec{F}_{\rm rot}+\epsilon^2\left(\vec{F}_{\rm EIH}+\vec{F}_{\rm SO}+\vec{F}_{\rm SS}+\vec{F}_{\rm B}+\vec{F}_{\rm I}+\Delta\vec{F}_{\rm B}\right)\;,
\end{equation}
where $\epsilon$ is a PN book-keeping parameter, $M_B$ and $\vec{a}_B$ are the relativistic mass and the orbital acceleration of the body B, $\vec{F}_N$ is the Newtonian force, $\vec{F}_{\rm tide}$ and $\vec{F}_{\rm rot}$ are perturbing forces caused by the tidal and rotational defomrmations of the body, $\vec{F}_{\rm EIH}$ is the Einstein-Infeld-Hoffmann force , $\vec{F}_{\rm SO}$ and $\vec{F}_{\rm SS}$ are the PN forces due to the spin-orbit and spin-spin coupling, $\vec{F}_{\rm B}$ is the PN force due to the second moment of inertia of the body, $\vec{F}_{\rm I}$ is the force due to the forth and higher-order moments of inertia, and $\Delta\vec{F}_{\rm B}$ is the PN force due to the second and higher order moments of inertia that exists only in the scalar-tensor theory of gravity. 

The PN forces are approximated as follows \cite{kv}:
\begin{equation}\label{4}  
{F}_{\rm EIH}\simeq  \left(\frac{v}{c}\right)^2F_N\;,
\end{equation}\begin{equation}\label{5}
{F}_{\rm SO}\simeq  \left(\frac{v}{c}\right)\left(\frac{v_{\rm r}}{c}\right)\left(\frac{L}{R}\right)F_N\;,\end{equation}\begin{equation}\label{6}
{F}_{\rm SS}\simeq  \left(\frac{v_{\rm r}}{c}\right)^2\left(\frac{L}{R}\right)^2F_N\;,\end{equation}\begin{equation}\label{7}
{F}_{\rm B}\simeq \kappa\left(\frac{v}{c}\right)^2\left(\frac{L}{R}\right)^2F_N\;,\end{equation}\begin{equation}\label{8} 
\Delta{F}_{\rm B}\simeq \kappa(\beta-1)\left(\frac{v}{c}\right)^2\left(\frac{L}{R}\right)^2F_N\;,\end{equation}\begin{equation}\label{9} 
{F}_{\rm I}\simeq \lambda\left(\frac{v}{c}\right)^2\left(\frac{L}{R}\right)^4F_N\;,
\end{equation}
where $\kappa$ and $\lambda$ are numerical factors depending on the internal distribution of density inside the stars, and $\beta$ is the non-linear gravity-coupling parameter of the PPN formalism \cite{will}. 

One can see that if stars have finite radius, there are the PN corrections to the EIH force that describe motion of point-like masses. The finite size PN forces (\ref{7})--(\ref{9}) are governed by the rotational moments of inertia which crucially depend on the internal structure of the stars even if they are spherically-symmetric. This property of the PN mechanics differs from the Newtonian mechanics of N-body problem.

PN forces $\vec{F}_{SO}$ and $\vec{F}_{SS}$ do not violate the effacing principle. We have proved \cite{kv} that the force $\vec{F}_{\rm B}$ can be completely eliminated from the equations of motion by choosing relativistic definition of the center of mass \cite{kv}. Hence, this force is not physical, and must be excluded from the theoretical analysis of the equations of motion. However, $\vec{F}_{\rm B}$ is to be retained for proper analysis of observations as the center of mass of each star is not known before the observations have been done, and must be considered as a fitting parameter \cite{km}.  

It is instructive to evaluate the limiting case of condensed astrophysical bodies like NS and BH. In this case, radius $L$ of the star is close to the Schwarzshild radius $R_g\sim 2GM/c^2$. We assume that BH is rotating with a limiting speed approaching $c$. Then, the forces (\ref{1})--(\ref{9}) are reduced to the following expressions
\begin{eqnarray}\label{10}  
F_{\rm tide}&\simeq& \kappa_{\rm tide}\left(\frac{v}{c}\right)^{10} F_N\;,
\qquad\qquad 
F_{\rm rot}\simeq \kappa_{\rm rot}\left(\frac{v}{c}\right)^{10} F_N\;,
\\\label{12}  
{F}_{\rm EIH}&\simeq&\left(\frac{v}{c}\right)^2F_N\;,
\quad\quad
{F}_{\rm SO}\simeq  \left(\frac{v}{c}\right)^3F_N\;,\quad\quad
{F}_{\rm SS}\simeq  \left(\frac{L}{R}\right)^4F_N\;,\\\label{15}
{F}_{\rm B}&\simeq& \kappa\left(\frac{v}{c}\right)^6F_N\;,\quad\quad
\Delta{F}_{\rm B}\simeq (\beta-1)F_B\;,\quad\quad
{F}_{\rm I}\simeq \lambda\left(\frac{v}{c}\right)^{10}F_N\;.
\end{eqnarray}
One can see that for the condensed astrophysical objects the effacing principle is violated in general relativity only in terms of the 5-th PN order. In scalar-tensor theory of gravity ($\beta\not=1$) this violation is of the 3-d PN order.

\vfill

\begin{thebibliography}{00}
\bibitem{dam} Damour, T.\ 1983, in: {\it Gravitational 
Radiation} (Amsterdam: North-Holland), pp. 59--144 
\bibitem{kop85} Kopeikin, S.~M.\ 1985, {\it Sov. Astron.}, {\bf 29}, 516 
\bibitem{alex}Alexander, M.~E.\ 1973, {\it Astrophys. Space Sci.}, {\bf 23}, 459 
\bibitem{kv} Kopeikin, S. \& Vlasov, I.\ 2004, {\it Phys. Rep.}, {\bf 400}, 209 
\bibitem{will} Will, C.~M. \& Nordtvedt, K.~J.\ 1972, {\it Astrophys. J.}, {\bf 177}, 757 
\bibitem{km}Kopeikin, S. \& Makarov, V.\ 2006, arXiv: astro-ph/0611358 






\end{thebibliography}
\end{document}